# Superconductivity near 80 Kelvin in single crystals of La$_3$Ni$_2$O$_7$ under pressure

Hualei Sun[1,7], Mengwu Huo[1,7], Xunwu Hu[1], Jingyuan Li[1], Yifeng Han[2], Lingyun Tang[3], Zhongquan Mao[3], Pengtao Yang[4], Bosen Wang[4], Jinguang Cheng[4], Dao-Xin Yao[1], Guang-Ming Zhang[5,6,*] & Meng Wang[1,*]

[1]Center for Neutron Science and Technology, Guangdong Provincial Key Laboratory of Magnetoelectric Physics and Devices, School of Physics, Sun Yat-Sen University, Guangzhou, Guangdong 510275, China

[2]School of Molecular Sciences and Center for Materials of the Universe, Arizona State University, Tempe, Arizona 85287, USA

[3]School of Physics and Optoelectronics, South China University of Technology, Guangzhou, Guangdong 510641, China

[4]Beijing National Laboratory for Condensed Matter Physics and Institute of Physics, Chinese Academy of Sciences, Beijing 100190, China

[5]State Key Laboratory for Low dimensional Quantum Physics, Department of Physics, Tsinghua University, Beijing 100084, China

[6]Collaborative Innovation Center of Quantum Matter, Beijing 100084, China.

[7]These authors contributed equally to this work.

*e-mail: wangmeng5@mail.sysu.edu.cn; gmzhang@mail.tsinghua.edu.cn

**Abstract**

High-transition-temperature (high-$T_c$) superconductivity in cuprates has been discovered for more than three decades, but the underlying mechanism remains a mystery. Cuprates are the only unconventional superconducting family that host bulk superconductivity with $T_c$s above the liquid nitrogen boiling temperature at 77 Kelvin. Here we report an observation of superconductivity in single crystals of La$_3$Ni$_2$O$_7$ with a maximum $T_c$ of 80 Kelvin at pressures between $14.0 - 43.5$ gigapascals using high-pressure resistance and mutual inductive magnetic susceptibility measurements. The superconducting phase under high pressure exhibits an orthorhombic structure of *Fmmm* space group with the $3d_{x^2-y^2}$ and $3d_{z^2}$ orbitals of Ni cations strongly mixing with oxygen *2p* orbitals. Our density functional theory calculations suggest the superconductivity emerges coincidently with the metallization of the *σ*-bonding bands under the Fermi level, consisting of the $3d_{z^2}$ orbitals with the apical oxygens connecting Ni-O bilayers. Thus, our discoveries not only reveal important clues for the high-$T_c$ superconductivity in this Ruddlesden-Popper double-layered perovskite nickelates but also provide a new family of compounds to investigate the high-$T_c$ superconductivity mechanism.

**Introduction**

The high-$T_c$ superconductivity in cuprates emerges from hole carriers doped to the Mott insulating state with a half-filled Cu $3d^9$ electronic configuration and $S = 1/2$ spin state[1-4]. As carriers doping and binging of the intra-layer Cu-O electronic *σ*-bonding, the so-called Zhang-Rice singlet forms, leading to the high-$T_c$ superconducting phase[5]. Around the optimal doping, it has been established that superconductivity has a *d*-wave pairing with gap nodes around the Brillouin zone diagonals[6-8]. Layered structure consisting of corner-connected CuO$_6$ octahedra and *Ln*O (*Ln* = lanthanide) layers is a common feature for the high-$T_c$ superconducting materials. Extensive efforts have been devoted to searching for superconductivity in nickel oxide compounds akin to cuprates[9-11]. Infinite-layer nickelates are one of the extensively investigated families, where Ni$^+$ ($3d^9$) shows the same electronic configuration as Cu$^{2+}$ cations. Until recently, Li, et. al. indeed found superconductivity with $T_c$ around $9 - 15$ K in the

Nd$_{0.8}$Sr$_{0.2}$NiO$_2$ thin films[12]. Afterward, superconductivity was observed in other hole-doped $Ln$NiO$_2$ thin films with infinite NiO$_2$ layers and Nd$_6$Ni$_5$O$_{12}$ with quintuple NiO$_2$ layers[13,14]. The maximum $T_c$ of 31 K has been achieved in Pr$_{0.82}$Sr$_{0.18}$NiO$_2$ films under 12.1 GPa, still below the so-called McMillan limit at 40 K[15]. Superconductivity all appears in the reduced Ruddlesden-Popper (RP) phases with the chemical formula $Ln_{n+1}Ni_nO_{2n+2}$, which are obtained from the RP phase $Ln_{n+1}Ni_nO_{3n+1}$ through removing two apical oxygens by a topochemical reduction method. A recent study suggests the unavoidable hydrogen in nickelate films is crucial for superconductivity[16]. On the other hand, no progress has been made on the observation of superconductivity in the RP phase or bulk samples of nickelates[17-19].

Among the RP phase nickelates, the trilayer square planar NiO$_2$ compounds attracted more attention, because the valence state of Ni cations in the reduced RP phase is +1.33, close to +1.2, where the maximum $T_c$ is expected theoretically[20,21]. In this work, we focus on the bilayer RP bulk single crystals of La$_3$Ni$_2$O$_7$[22]. A simple electron count gives a Ni$^{2.5+}$, i.e., $3d^{7.5}$ state for both Ni cations, and experiments indicate that La$_3$Ni$_2$O$_7$ is a paramagnetic metal[23]. Ni$^{2.5+}$ is usually believed to be given by mixed valence states of Ni$^{2+}$ ($3d^8$) and Ni$^{3+}$ ($3d^7$), corresponding to the half-filled of both $3d_{z^2}$ and $3d_{x^2-y^2}$ orbitals and single occupied $3d_{z^2}$ with empty $3d_{x^2-y^2}$ orbitals, respectively. For the two nearest intra-layer Ni cations in a bilayer RP phase, however, two $3d_{z^2}$ orbitals via apical oxygen usually have a large inter-layer coupling due to the quantum confinement of the NiO$_2$ bilayer in the structure, and the resulting energy splitting of Ni cations can dramatically change the distribution of the averaged valence state of +2.5. First, we have synthesized La$_3$Ni$_2$O$_7$ single crystals using a high-pressure floating zone method. The structure of La$_3$Ni$_2$O$_7$ crystallizes into an orthorhombic phase (space group *Amam*), with a corner-connected NiO$_6$ octahedral layer separated by a La-O fluorite-type layer stacking along the *c* axis[22,23], as shown in Fig. 1. Then the structure was investigated at pressures up to 41.2 GPa, as applying pressure is effective to induce the Jahn-Teller effect through structural modification and the electronic band structures. The synchrotron X-ray diffraction (XRD) patterns in the low-pressure (LP) phase from 1.6 to 10.0 GPa can be well indexed by the orthorhombic *Amam* space group (Fig. 1a). An anomaly in the positions of the reflection peaks occurs around 10 GPa, suggesting a structural transition (Fig. 1b). Our density functional theory (DFT) calculations suggest the structure transforms from the *Amam* to the *Fmmm* space group under pressure (Fig. 1c). The XRD patterns in the high-pressure (HP) phase above 15.0 GPa can be indexed by the orthorhombic *Fmmm* space group. Evolutions of the lattice parameters and unit cell volume as a function of pressure confirm the structural transition and the DFT calculations (Fig. 1c and 1d). In particular, the space group transition corresponds to the change of the bond angle of Ni-O-Ni from 168.0° to 180° along the *c* axis, as depicted in Fig. 1f. We notice that X-ray is not sensitive to the position and content of oxygens. The structure of the HP phase is determined by a combination of the DFT calculations and the XRD refinements. The structural parameters of La$_3$Ni$_2$O$_7$ refined at 1.6 GPa and 29.5 GPa are listed in Extended Data Table 1. We notice that the inter-atomic distance between the Ni and apical oxygen is abruptly reduced from 2.297 Å in the *Amam* phase at ambient to 2.122 Å in the *Fmmm* structure at 32.5 GPa (Extended Data Fig. 1).

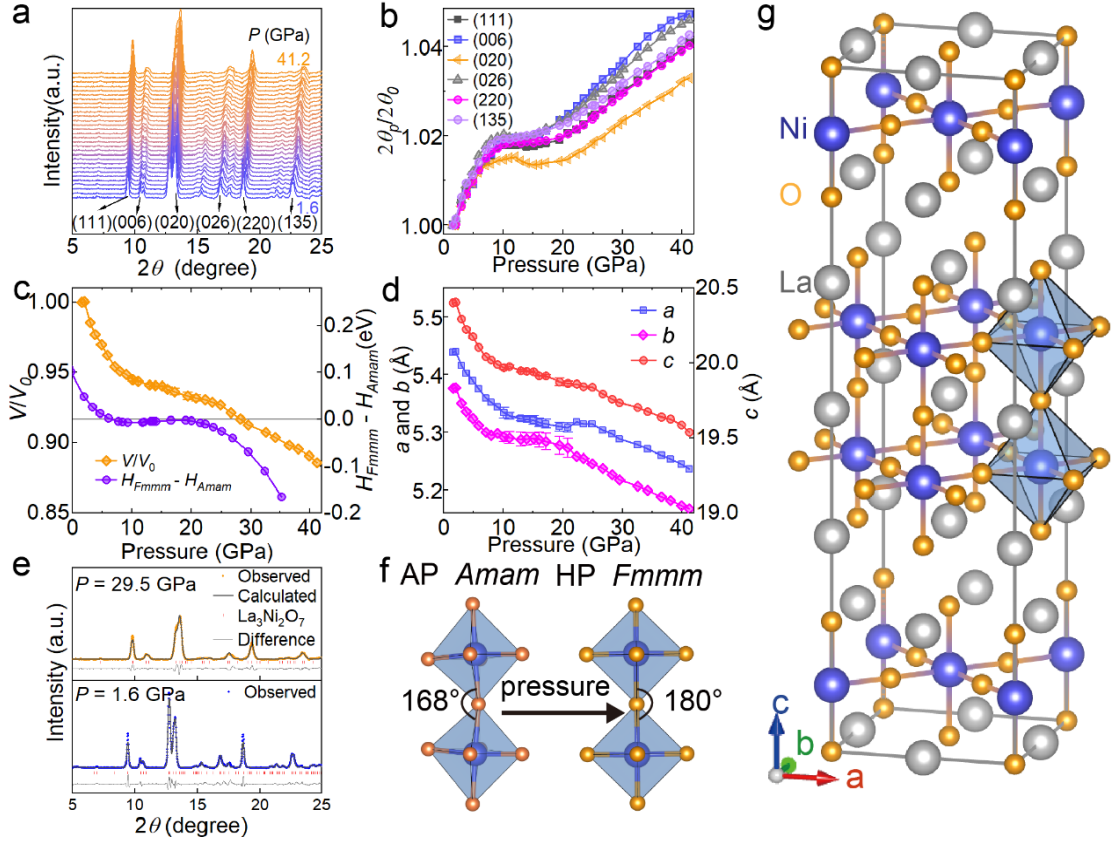

**Fig. 1 Structural characterizations of pressurized La$_3$Ni$_2$O$_7$. a**, Synchrotron X-ray diffraction patterns of powder samples under various pressures between 1.6 and 41.2 GPa. **b**, Pressure dependence of the peak positions labeled by the Miller indexes of the *Amam* space group at ambient pressure. **c**, The orange diamonds represent the change of volume as a function of pressure determined from experiments. The error bars for pressures between 10 − 20 GPa are determined from the differences of the volumes refined as the *Amam* and *Fmmm* space groups, respectively. The violet circles represent the difference of the enthalpy for one cell between the space group *Fmmm* and *Amam* as a function of pressure calculated as the first principles method. The enthalpy is defined as $H = E(V) + PV$, where $E(V)$ is energy. The results reveal the ground structure changes to the *Fmmm* space group under high pressures. **d**, Lattice constants *a*, *b*, and *c* refined from the X-ray diffraction patterns. **e**, Refinements of the synchrotron X-ray diffraction patterns at 29.5 GPa (the upper panel) using the space group *Fmmm* and 1.6 GPa (the lower panel) using the space group *Amam*. **f**, The Ni-O-Ni angle between two adjacent octahedra as shaded in cyan changes from 168° in the ambient pressure *Amam* space group to 180° in the high-pressure *Fmmm* space group. **g**, Crystal structure of La$_3$Ni$_2$O$_7$ with the orthorhombic structure.

To elucidate the electronic structure of La$_3$Ni$_2$O$_7$ under pressure, we conducted DFT calculations at 1.6 and 29.5 GPa, respectively. The electronic structure can be understood by the crystal field splitting of the NiO$_6$ octahedron on the $e_g$ and $t_{2g}$ orbitals of Ni cations[24]. Results of the non-magnetic solution at 1.6 GPa reveal that, the electronic states of Ni $3d_{x^2-y^2}$ and $3d_{z^2}$ orbitals are well separated from other three Ni $t_{2g}$ orbitals in the energy range of (-2 eV, 2 eV), and the Ni $3d_{x^2-y^2}$ orbitals with oxygen $2p$ orbitals dominating across the Fermi level (Fig. 2a), reminiscent that of cuprates[25]. The sizes of the hole Fermi surfaces around Γ and the electron Fermi surfaces around Y are comparable. Below and above the Fermi level, there exists electronic bonding and anti-bonding bands of the $3d_{z^2}$ electronic states due to the large inter-layer $\sigma$-bond coupling through an inner apical oxygen (Fig. 2d). There appears a band gap due to the quantum confinement of the NiO$_2$ bilayer in the structure. The splitting makes the $3d_{z^2}$ bonding bands lower in energy and fully occupied, while the $3d_{x^2-y^2}$ bands are still degenerate and have a quarter-filling (Fig. 2e). Moreover, the bonding $3d_{z^2}$ electronic states form rather flat bands along both the $\Gamma - X$ and $\Gamma - Y$ directions, which are the main characteristics of the electronic $\sigma$-bonds. For the HP phase at 29.5 GPa, however, the $3d_{z^2}$ bonding bands lift upward crossing the Fermi level as the apical oxygens are hole doped (Fig. 2b) and a small hole Fermi pocket emerge around the center of the Brillouin zone (Extended Data Fig. 2), corresponding to the metallization of the lower $\sigma$-bonds precisely. Meanwhile, the same number of electrons is added to the Ni $3d_{x^2-y^2}$ orbitals, increasing their electron occupation evidenced by the wider proportion of $3d_{x^2-y^2}$ bands below the Fermi level (Fig. 2b). The occupied electrons of Ni $3d_{x^2-y^2}$ orbitals are expected to strongly interact with the $2p$ orbital of oxygens, forming the intra-layer Zhang-Rice singlets[5]. The obtained total densities of states reach to a maximum at the Fermi energy (Fig. 2b). As high pressure is applied, the electronic interactions between the bilayers of NiO$_2$ are increased, evidenced by the enlarged splitting of the $3d_{z^2}$ orbitals at the X and Y points. Therefore, the unique feature of the pressurized electronic structure makes the filling of electrons for the two Ni$^{2.5+}$ cations resemble that of Cu$^{2+}$ ($3d^9$) in hole doped bilayer cuprates[26]. The metallization of the inter-layer $\sigma$-bonding bands recalls the emergence of the conventional high-$T_c$ superconductivity in MgB$_2$, Li$_3$B$_4$C$_2$, H$_3$S and other H-enriched compounds[27-29], motivating us to explore the possible superconductivity in the HP phase of La$_3$Ni$_2$O$_7$.

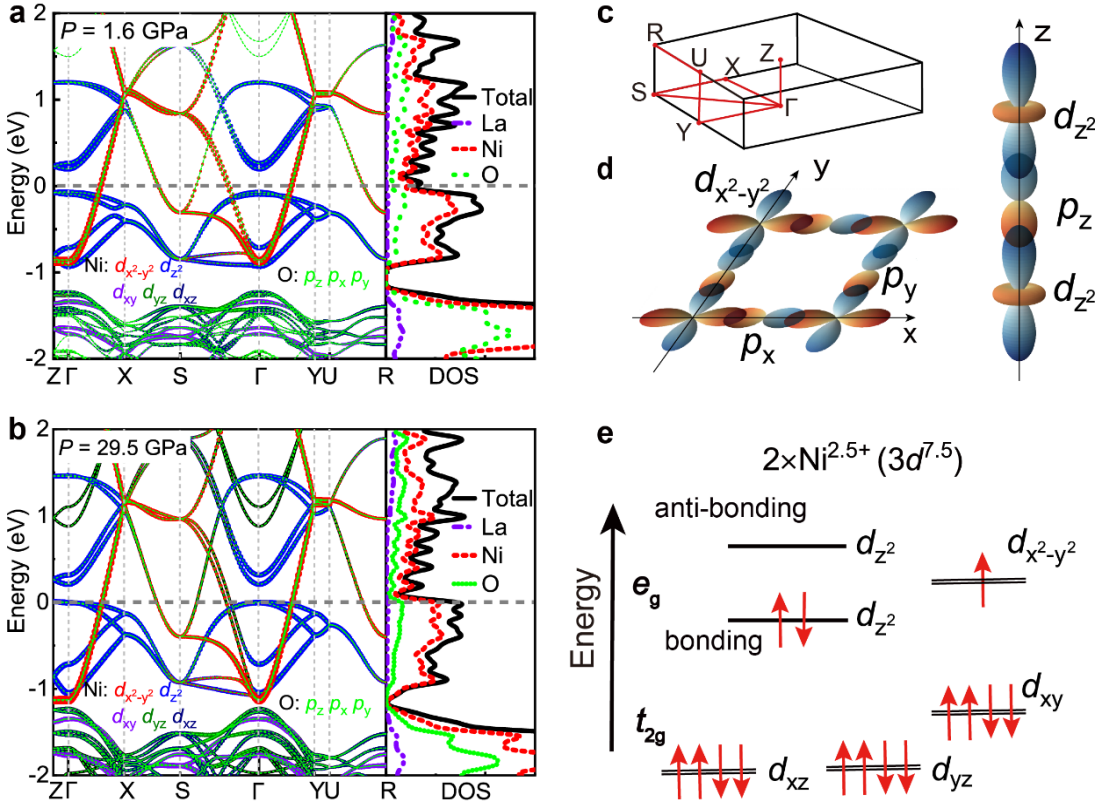

**Fig. 2 Density functional theory calculations for La$_3$Ni$_2$O$_7$ at 1.6 and 29.5 GPa. a,** Projected electronic band structures of Ni cations and O anions in La$_3$Ni$_2$O$_7$ calculated using the structural parameters relaxed from the synchrotron X-ray diffraction at 1.6 GPa and a Coulomb repulsive $U = 4$ eV. The right is the corresponding density of states near the Fermi level. The violet, red, and green curves

represent the contributions from La, Ni, and O, respectively. **b**, Projected electronic band structures and the density of states at 29.5 GPa. **c**, Schematic of the three-dimensional orthorhombic Brillouin zone. The red lines correspond to the paths of the electronic bands in **a** and **b**. **d**, Schematic of the intralayer σ-bonding states formed by the strong interaction of the Ni $3d_{x^2-y^2}$ and O $2p_{x/y}$ orbitals (left). The interlayer σ-bonding-antibonding states consisting of the Ni $3d_{z^2}$ and O $2p_z$ orbitals (right). **e**, Electronic configuration of two $Ni^{2.5+}$ ($3d^{7.5}$) in the environment of bilayers of $NiO_6$ octahedra. The $d_{z^2}$ orbitals of two Ni cations in the adjacent layers split due to the bonding and anti-bonding states.

Figure 3a shows the temperature dependence of the resistance for $La_3Ni_2O_7$ single crystals in the pressure range of 0 − 18.5 GPa. At ambient pressure, $La_3Ni_2O_7$ is metallic and behaviors like a Fermi liquid. The single crystals used for high pressure measurements were taken from the sample measured at ambient. A pressure of 1.0 GPa can change the ground state from metallic to weakly insulating, consistent with previous reports[30,31]. The increase of resistance under a small pressure could be ascribed to the distortion of the $NiO_6$ octahedra[32]. With further increasing pressure, $La_3Ni_2O_7$ undergoes a weakly insulating to metallic transition at ~10 GPa, and a clear drop in resistance at ~78.2 K appears at pressures above 14.0 GPa, indicating a superconducting-like phase transition. The temperatures of the drop are weakly pressure dependent, reaching to 80 K at 18.5 GPa. Above the transition temperature, the resistance increases linearly up to 300 K, which is a typical property of a strange metal state, characterizing the normal state of the optimal doped cuprate superconductors[33]. Since pressure gradient and internal strain effect could affect the electric transport properties under pressure, as observed in $K_{0.8}Fe_{1.7}Se_2$ (Ref.34) and $BaFe_2S_3$ (Ref.35), we employed a soft material KBr as the pressure-transmitting medium and the resistance is measured under higher pressures (Fig. 3b). The sharp drops on resistance and the flat resistance that closes to zero below the transition temperature ($T_c$s) suggest a superconducting transition. The nonzero resistance may be related to the pressure-transmitting medium. In $BaFe_2S_3$, the zero resistance is achieved by using the liquid transmitting medium - glycerine[35], while the present resistance for different single crystal samples of $La_3Ni_2O_7$ under pressure is measured without a pressure-transmitting medium. The behaviors of resistance are repeatable, as shown in Extended Data Fig. 3.

To show the diamagnetic property of our samples under high pressure, we measured the inductive voltage, which can be regarded as alternating current (ac) magnetic susceptibility[36], for the single crystal $La_3Ni_2O_7$ under pressures up to 28.7 GPa by using a diamond anvil cell and the mutual induction method[37]. There is an obvious diamagnetic response below 77 K at 25.2 GPa from the real part of the ac magnetic susceptibility χ'(T), as shown in Fig. 3c. The measured electronic and magnetic properties demonstrate that the transition near 80 K corresponds to the emergence of superconductivity. Fig. 3d shows the evolution of the resistance at 18.9 GPa under various magnetic fields up to 14 T. The field-suppressed superconductivity is more pronounced at lower temperatures. This is analogous to the cuprate superconductors, where the onset $T_c$ is kept unchanged[38]. The upper critical field $\mu_0 H_{C2}(0)$ of $La_3Ni_2O_7$ has been determined using the criterion of 0.9×$R(T_c^{onset})$, where $R(T_c^{onset})$ is the resistance at the onset $T_c$. The Ginzburg–Landau formula is adopted for fitting $\mu_0 H_{C2}$s for various pressures, yielding the highest $\mu_0 H_{C2}$ = 186 T for 18.9 GPa, as shown in Fig. 3e. An estimation of the in-plane superconducting coherence length is 4.83 nm for 18.9 GPa at zero temperature.

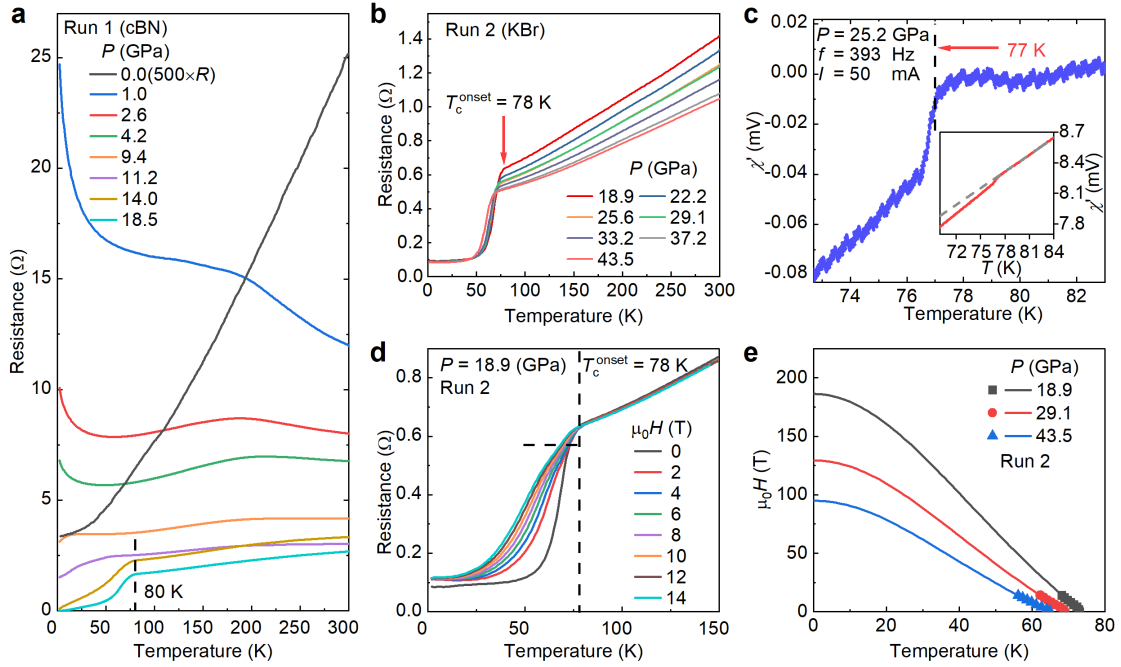

**Fig. 3. Superconducting transitions in $La_3Ni_2O_7$ single crystals under pressure. a**, Resistance of $La_3Ni_2O_7$ versus temperature at different pressures from 1.0 to 18.5 GPa with the gasket of cubic boron nitride (cBN) and epoxy mixture. The resistance at ambient pressure was measured independently. **b**, High pressure resistance measurements using KBr as the pressure-transmitting medium. The arrow shows the onset superconducting transition temperatures ($T_c$s). The onset $T_c$ at 18.9 GPa is 78 K. **c**, The background-subtracted real part of the ac susceptibility showing a prominent diamagnetic response at 25.2 GPa with a current frequency of 393 Hz and magnitude of 50 mA. The red solid line in the inset is the raw data and the gray dashed line is a fitted background following the trend above the transition at 77 K. The vertical dashed line marks the $T_c$. **d**, Resistance curves below 150 K at different magnetic fields from 0 to 14 T at 18.9 GPa. The vertical dashed lines show the onset $T_c$s. The horizontal line in **d** indicates 90% of the resistance at the onset $T_c$, $0.9 \times R(T_c^{onset})$. The currents used in **a**, **b**, and **d** are 10, 300, and 300 $\mu A$, respectively. **e**, The Ginzburg-Landau fittings of the upper critical fields, $\mu_0 H_{C2}$ at pressures of 18.9, 29.1, and 43.4 GPa by using the criteria of the $0.9 \times R(T_c^{onset})$ for the Run 2. The maximum $\mu_0 H_{C2}$ for 18.9 GPa is 186 T. The magnetic fields are applied along the $c$ direction.

The electrical and magnetic measurements under high pressure were repeated on several single-crystal samples (Extended Data Fig. 3, Fig. 4, and Fig. 5). The corresponding $T_c$s are summarized in the temperature-pressure phase diagram in Fig. 4. The transition from a weak insulating phase to a superconducting phase against pressure is analogous to the hole-doping dependence of superconductivity in the infinite-layer nickelate films[16,39-41]. The superconductivity with a high transition temperature above the liquid nitrogen boiling point at 80 K emerges in the orthorhombic $Fmmm$ phase. However, the $T_c$ values are not dramatically changed in the superconducting region. The normal state of the superconductivity shows a strange metal behavior that is characterized by a linear temperature-dependent resistance up to 300 K. This may indicate that the HP metallic phase is close to a quantum critical regime adjacent to the superconducting phase.

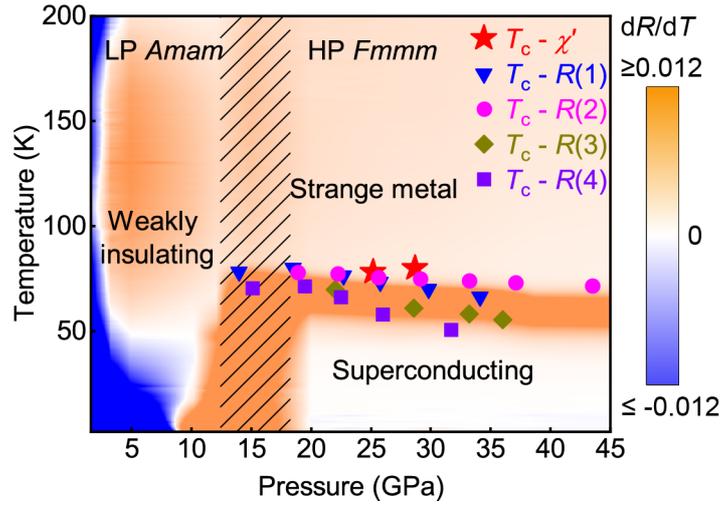

**Fig. 4 Phase diagram of the high-temperature superconductivity in La$_3$Ni$_2$O$_7$ single crystals.** The superconducting transition temperatures, $T_c$s, as a function of pressure obtained from our resistance and inductive magnetic susceptibility measurements are depicted. The colors of the background refer to the derivative of the resistance with respect to temperature in Figs. 3a and 3b. The blue color with negative values of $dR/dT$ indicates the decreasing of resistivity as increasing temperature. The homogenous color against temperature in the strange metal area stands for the linear temperature dependent resistance. The shaded stripe indicates a structural transition from the low-pressure orthorhombic *Amam* phase to the high-pressure *Fmmm* phase.

It should be mentioned that some samples are insulating at low pressures and cannot be tuned to metallic up to 20 GPa. These should be related to the presence of slightly deficient oxygens in La$_3$Ni$_2$O$_{7-\delta}$. Previous experiments[42,43] have showed that, when the oxygen-deficient variant $0.08 \leqslant \delta \leqslant 0.63$, a structure phase transition occurs from orthorhombic to tetragonal symmetry and the electrical transport properties change from metallic to weakly insulating at ambient pressure. Concerning with the metallic behavior of our superconducting samples, the deficient oxygen variant δ should be less than 0.08.

In summary, we have shown that the electronic occupancy of Ni$^{2.5+}$ in the RP double-layered perovskite nickelate La$_3$Ni$_2$O$_7$ can be mimic to the Cu$^{2+}$ of hole doped bilayer high-$T_c$ cuprates due to the presence of strong inter-layer coupling of $3d_{z^2}$ orbitals via the apical oxygen anions. Such an inter-layer coupling results in the formation of the interlayer σ-bonding and anti-bonding bands lying below and above the Fermi level. Applying a high pressure can realize the metallization of the σ-bonding bands below the Fermi level through the hole doping in the $3d_{z^2}$ orbitals and electron doping in the $3d_{x^2-y^2}$ orbitals. These are the most important clues of the high-$T_c$ superconductivity with $T_c \sim 80$ K observed in La$_3$Ni$_2$O$_7$ single crystals under pressure. Although both $3d_{x^2-y^2}$ and $3d_{z^2}$ orbitals are involved, those features are distinctly different from the infinite-layer superconducting nickelates, where the electronic states of oxygen *2p* orbitals are far below the Fermi level and have a much reduced *3d-2p* mixing due to the larger separation of their site energies. From our experiments, the $T_c$ is comparable with that of the high-$T_c$ cuprate superconductors[1] and higher than the record $T_c$ of the iron-based superconductors[4,44]. To our knowledge, this is the first experimental report on the superconductivity in bulk nickelates and the first discovery of superconductivity in the RP phase of nickelates. Our results demonstrate that the nickel oxide system is a fascinating and plausible platform to search for high-$T_c$ superconductors and explore its unconventional high-$T_c$ superconductivity mechanism.

## Methods

**Material synthesis.** La$_3$Ni$_2$O$_7$ single crystals were grown by a vertical optical-image floating zone furnace with an oxygen pressure of 15 bar and a 5 kW Xenon arc lamp (100-bar Model HKZ, SciDre GmbH, Dresden)[23].

**High-pressure synchrotron X-ray experiments.** High-pressure synchrotron radiation X-ray diffraction data were collected at 300 K with the wavelength λ = 0.6199 Å at Beijing Synchrotron Radiation Facility. An asymmetric diamond anvil cell with a pair of 300-μm-diameter culets was used. The steel gasket was pre-indented and a diameter of 110 μm was laser-drilled at the center to serve as a sample chamber. The samples were ground into powder and a ruby sphere was loaded into the middle of the sample chamber and silicone oil was used as a pressure-transmitting medium. The pressure was calibrated by measuring the shift of its fluorescence wavelength. The data were initially integrated using Dioptas (with a CeO$_2$ calibration)[45] and the subsequent Rietveld refinements were processed using TOPAS-Academic[46].

**High-pressure magnetic and electrical property measurements.** The electrical resistance measurements of La$_3$Ni$_2$O$_7$ single crystals were performed using the standard four-probe method. High pressures were generated with screw-pressure-type diamond anvil cells made of nonmagnetic Be-Cu alloy. Diamond anvils with a 400 μm culet were used and the corresponding sample chamber with a diameter of 150 μm was made in an insulating gasket achieved by cubic boron nitride and epoxy mixture. A single crystal with dimension of 80 × 60 × 10 μm$^3$ was loaded without pressure-transmitting medium in run 1. Fine KBr powders as the pressure-transmitting medium and a single crystal basically with the same size were adopted in run 2. Pressure was calibrated by using the ruby fluorescence shift at room temperature for all experiments. Electrical measurements were taken on a physical property measurement system (PPMS, Quantum Design) providing synergetic extreme environments with temperatures from 2 to 300 K and magnetic fields up to 14 T.

**Magnetic susceptibility measurements.** Alternating-current (*ac*) magnetic susceptibility up to 14.3 GPa using a palm-type cubic anvil cell (CAC) was measured at the Synergetic Extreme Condition User Facility (SECUF). For these measurements with CAC, no discernable anomaly can be detected in the studied pressure range (below 14.3 GPa). Then, we measured the *ac* magnetic susceptibility up to 28.7 GPa by using a magnetic inductive technique[34,36] in the School of Physics and Optoelectronics, South China University of Technology. The 600-μm diamond culets and the corresponding sample chamber with a diameter of 180 μm were made in a nonmagnetic Be-Cu gasket. The sample chamber was filled with fine La$_3$Ni$_2$O$_7$ powder without any other pressure-transmitting medium. Pressure values were estimated from the calibration curve determined by the ruby fluorescence wavelength at 300 K. This magnetic inductive technique consists of three parts, including exciting coil, pickup coil, and compensating coil. The alternating current in the exciting coil of 100 turns with a diameter of 8.5 mm is fed from a Stanford Research SR830 digital lock-in amplifier. The corresponding excitation field is about 9 Oe. Inside the excitation coil, a pickup coil of 100 turns with a diameter of 2.0 mm is wounded around the sample, and a compensating coil is oppositely connected next to it. The alternating magnetic field generates electromotive forces in the pickup coil, which is detected by a Keithley 2182A. The detected signal is a superposition of the susceptibility of the metallic parts of the diamond anvil cell (DAC) and the susceptibility of the sample.

**Density functional theory calculations.** The first-principles calculations are performed using the density functional theory as implemented in the Vienna *ab initio* simulation package (VASP)[47]. The projector augmented-wave (PAW) method[48] with a 600 eV plane-wave kinetic cutoff energy is employed.

The generalized gradient approximation of Perdew-Burke-Ernzerhof[49] form is used for exchange-correlation functional. A 19 × 19 × 5 *k*-points mesh is used for the self-consistent and Fermi surface calculations. The lattice parameters are fixed to the experimentally refined lattice constants obtained from X-ray diffraction, and the atomic positions are fully optimized until forces on each atom are less than 0.001 eV/ Å, and the energy convergence criterion is set to be $10^{-6}$ eV for the electronic self-consistent loop.

For the DFT + *U* treatment of Ni 3*d*-electrons in $La_3Ni_2O_7$, the *U* parameter is estimated to 5.9 eV using the linear response method. We tested the *U* values with 4, 5, and 6 eV, which give similar results. Finally, an effective Hubbard *U* for the 3*d* electrons of Ni cations is chosen as 4 eV in this work[50]. The energies as a function of volume ($E-V$) are calculated from the first-principles calculations. The volume of the unit cell is fixed by using the experimental lattice parameters, and the atomic positions are fully optimized. The Murnaghan equation is employed to fit the $E-V$ data. The pressure dependence of the enthalpy can be written as $H = E(V) + PV$. The ground state phase can be determined from the enthalpy.

**Acknowledgements** M. W. acknowledges the support by the National Natural Science Foundation of China (grant no. 12174454), Guangdong Basic and Applied Basic Research Funds (grant No. 2021B1515120015), and Guangdong Provincial Key Laboratory of Magnetoelectric Physics and Devices (grant No. 2022B1212010008). H. S. acknowledges the support of Guangzhou Basic and Applied Basic Research Funds (grant No. 202201011123). D. X. Y. is supported by NKRDPC-2022YFA1402802, NKRDPC-2018YFA0306001, NSFC-92165204, NSFC-11974432, and Shenzhen International Quantum Academy. P.T.Y, B.S.W. and J.G.C. are supported by the National Natural Science Foundation of China (grant Nos. 12025408, 11921004), the Beijing Natural Science Foundation (grant No. Z190008), the National Key R&D Program of China (grant No. 2021YFA1400200), and the Strategic Priority Research Program of CAS (grant No. XDB33000000). A portion of this work was carried out at the Synergetic Extreme Condition User Facility (SECUF). High-pressure synchrotron X-ray measurements were performed at 4W2 High Pressure Station, Beijing Synchrotron Radiation Facility (BSRF), which is supported by Chinese Academy of Sciences (Grant KJCX2-SW-N20, KJCX2-SW-N03).

**Author contributions** M. W. designed the project; H. S., M. H. and J. L. performed the resistance measurements under pressure; H. S.  performed the synchrotron X-ray diffraction measurements; H. S. and J. L. conducted the high-pressure susceptibility measurements with the support of L. T. and Z. M.; Magnetic susceptibility for pressures below 14 GPa (data not shown) were measured with the support of P. Y., B. W., and J. C.; H. S., Y. H., and M. H. conducted the structural analysis; D. X. Y. and X. H. performed the density functional theory calculations. G. Z. proposed a relevant physical picture to understand both the numerical and experimental results. M. W. and G. Z. wrote the paper with inputs from all coauthors.

**Competing interests** The authors declare no competing interests.

**Correspondence and requests for materials** should be addressed to Meng Wang and Guang-Ming Zhang.

**Extended Data Table 1**

**Table 1 Lattice parameters refined from experiments and optimized theoretically.** Refined lattice parameters, atomic coordinates of La$_3$Ni$_2$O$_7$ at 1.6 and 29.5 GPa. The values in the brackets are parameters optimized by the density functional theory method and adopted in the calculations.

| | The LP phase at 1.6 GPa, space group: *Amam* | | | |
|---|---|---|---|---|
| | $a$ = 5.4392 (8), $b$ = 5.3768 (8), and $c$ = 20.403 (4) Å, $\alpha = \beta = \gamma = 90°$, $R_{wp}$ = 9.0%, $R_p$ = 15.7% | | | |
| atom | $x$ | $y$ | $z$ | Occ. |
| Ni | -0.750 | 0.750 [0.748] | 0.401(1) [0.405] | 1 |
| La1 | -0.250 | 0.750 [0.758] | 0.318 (5) [0.321] | 1 |
| La2 | -0.750 | 0.250 [0.248] | 0.500 | 1 |
| O1 | -1.000 | 1.000 | 0.410 | 1 |
| O2 | -0.750 | 0.710 [0.719] | 0.500 | 1 |
| O3 | -0.500 | 0.500 | 0.400 [0.398] | 1 |
| O4 | -0.750 | 0.780 [0.781] | 0.300 [0.295] | 1 |
| | The HP- phase at 29.5 GPa, space group: *Fmmm* | | | |
| | $a$ = 5.289 (2) $b$ = 5.218 (2) and $c$ = 19.734 (5) Å, $\alpha = \beta = \gamma = 90°$, $R_{wp}$ = 12.8%, $R_p$ = 16.7% | | | |
| atom | $x$ | $y$ | $z$ | *Occ* |
| Ni | 0 | 0 | 0.089 (2) [0.096] | 1 |
| La1 | 0 | 0 | 0.316 (5) [0.321] | 1 |
| La2 | 0 | 0 | 0.50 | 1 |
| O1 | 0.250 | 0.250 | 0.096 [0.095] | 1 |
| O2 | 0 | 0 | 0.190 [0.204] | 1 |
| O3 | 0 | 0 | 0 | 1 |

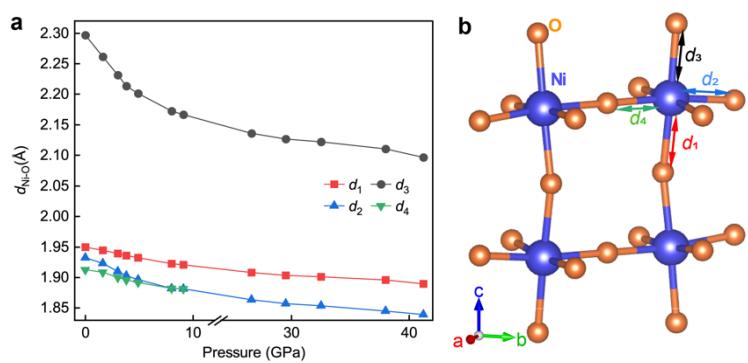

**Extended Data Fig. 1 Ni-O distances in the NiO$_6$ octahedra of La$_3$Ni$_2$O$_7$ under pressure. a,** Ni-O distances against pressure. The lattice constants are refined from synchrotron X-ray diffraction. The Ni-O distances are determined from optimization by the density functional theory and used in the calculations. **b,** Sketch of the NiO$_6$ octahedra. The $d_1$, $d_2$, $d_3$, and $d_4$ label the corresponding Ni-O distances.

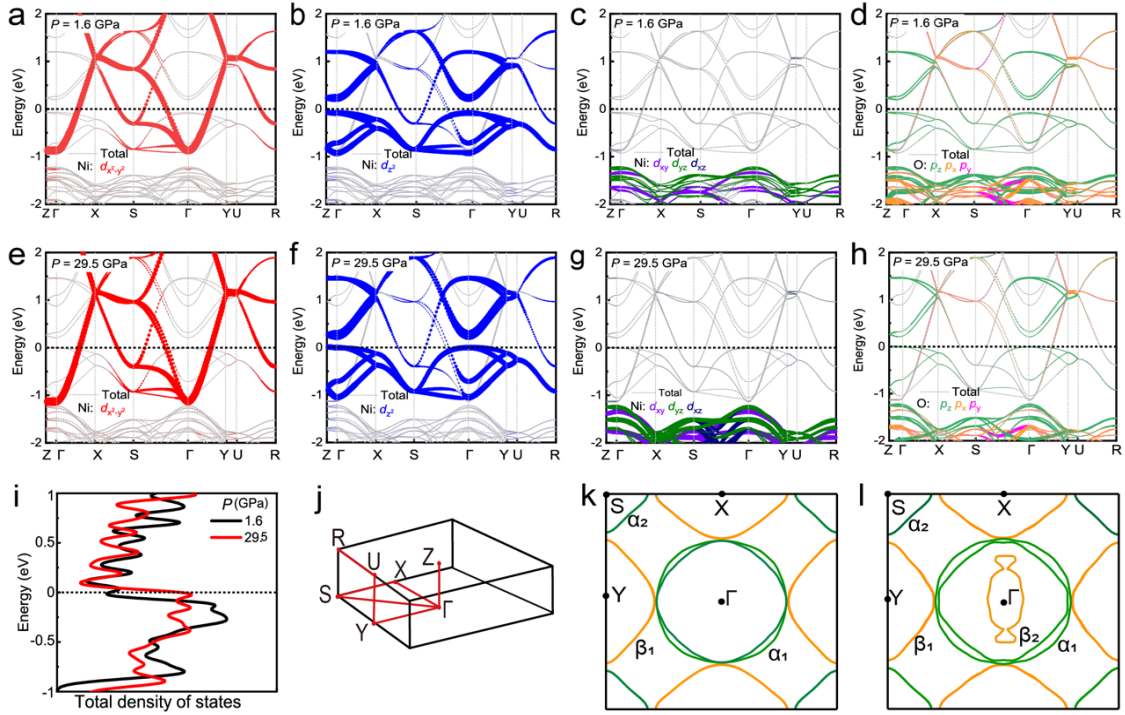

**Extended Data Fig. 2 Density functional theory calculations for La$_3$Ni$_2$O$_7$ at 1.6 and 29.5 GPa.** Orbital-decomposed band structures of La$_3$Ni$_2$O$_7$ at 1.6 GPa **(a)-(d)** and 29.5GPa **(e)-(f)**. **i**, The total density of states at 1.6 and 29.5 GPa near the Fermi level. **j**, Schematic of the three-dimensional reciprocal unit cell. The red lines correspond to the paths of the electronic bands. **k**, Calculated two-dimensional Fermi surfaces of La$_3$Ni$_2$O$_7$ in a Brillouin zone at 1.6 GPa marked by a black square. The Fermi surfaces consist of electrons bands ($\alpha_{1,2}$) and a hole band ($\beta_1$). **l**, Two-dimensional Fermi surfaces of La$_3$Ni$_2$O$_7$ at 29.5 GPa. Additional hole bands (Ni $3d_{z^2}$) cross the Fermi level.

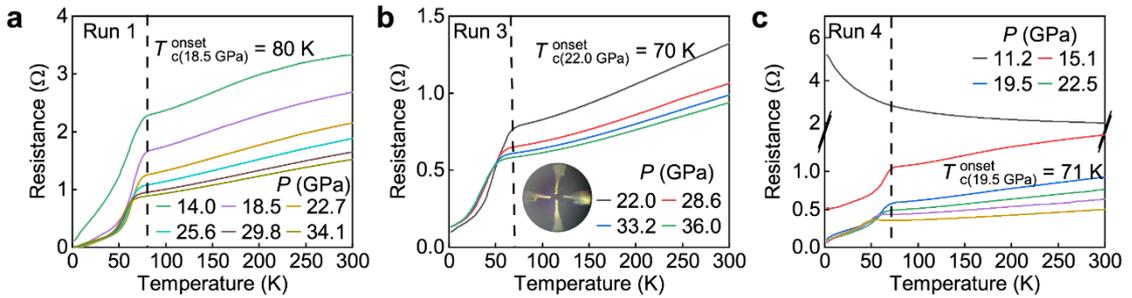

**Extended Data Fig. 3 Resistance measurements of La$_3$Ni$_2$O$_7$ single crystals under pressure acquired in different runs.** Resistance curves obtained from (**a**) Run 1, (**b**) Run 3, and (**b**) Run 4 measured with a gasket of cubic boron nitride without a pressure-transmitting medium. The vertical dashed lines indicate the onset superconducting transition temperature $T_c$. The inset in **a** is a photo showing the electrodes for the high-pressure measurements. A current of 10 $\mu$A was used for the measurements.

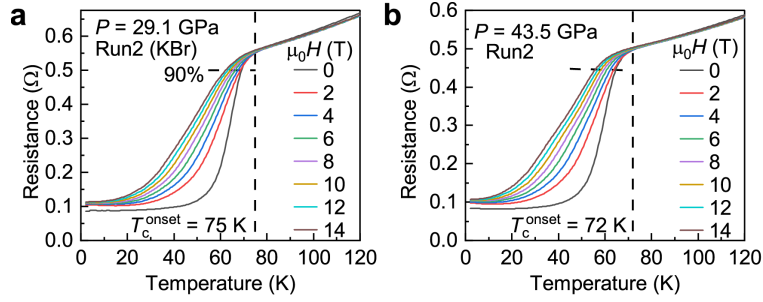

**Extended Data Fig. 4 Suppression of superconductivity of La$_3$Ni$_2$O$_7$ by external magnetic fields.** Resistance measured at (**a**) 29.1 GPa and (**b**) 43.5 GPa in the Run 2 with KBr as the pressure transmitting medium. The horizontal dashed lines mark $0.9 \times R(T_c^{onset})$, where $R(T_c^{onset})$ is the resistance at the onset $T_c$.

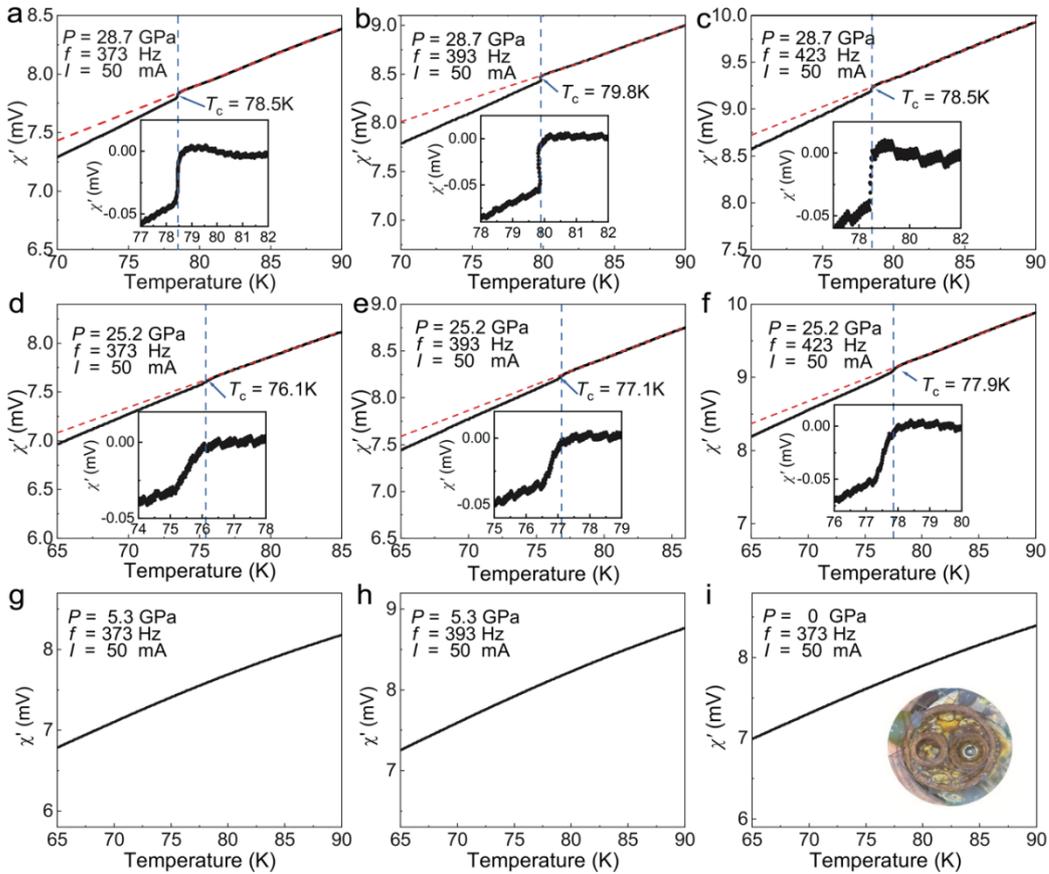

**Extended Data Fig. 5 Diamagnetic response measurements of La$_3$Ni$_2$O$_7$ under pressure.** Raw data of the real part of the ac susceptibility showing a prominent diamagnetic response at 28.7 GPa with a current magnitude of 50 mA and frequency of (**a**) 373, (**b**) 393, (**c**) 423 Hz and (**d-f**) identical measurements at 25.2 GPa. The red dashed lines are fitted backgrounds following the trend above the superconducting transitions. Insets in **a-f** show the diamagnetic signals obtained by subtracting the fitted linear backgrounds. The transition temperature shifts because the pressure changes for each measurement. **g** and **h**, Diamagnetic response measurements at 5.3 GPa measured during the decompressing process with a 373 and 393 Hz frequency current, respectively. **k**, The background measurement of the diamagnetic response of the cell without a sample. The inset in **k** is an image of the experimental set-up for the *ac* susceptibility measurements in a diamond-anvil cell, with a signal coil around the diamond anvils and a neighbor compensating coil.